# Challenges with the Application of Cyber Security for Airworthiness (CSA) in Real-World Contexts

(Beckett LeClair, James McLeod, Lee Ramsay and Mick Warren, 2023)
Frazer-Nash Consultancy

**Abstract**

The ever-increasing push towards reliance upon computerised technology in commercial, general, and military aerospace brings with it an increasing amount of potential cyber hazards and attacks; consequently, the variety of attack vectors is greater than ever. Recognized Good Practice (RGP) standards such as DO-326A[1] and ED-202A[2] attempt to address this by providing guidelines for cyber security on in-service aircraft, though implementation work for such initiatives is still in early stages. From previous work on in-service aircraft, the authors have determined that one of the key challenges is that of the retrospective application of new regulations to existing designs; this can present significant requirements for time, money, and Suitably Qualified and Experienced Personnel (SQEP) resource – things which are often in already limited supply in military environments. The authors have previously explored efficient ways of approaching compliance, with promising results. There is still the need to consider this retroactivity challenge in tandem with other key factors affecting the application of CSA[3], in order to determine any more potential mitigating actions that could lower the barrier to effective and efficient implementation of secure approaches in the air domain. This work explores the interrelated challenges surrounding real-world applications of CSA and the beginnings of how these may be overcome.

## 1 - Introduction

*1.1 What is Cybersecurity for Airworthiness?*

Traditionally, cybersecurity standards are concerned with protecting assets from intentional manipulation by threat actors. Conversely, safety standards are concerned with ensuring that safety risks arising from asset use are 'as low as reasonably practicable' (ALARP), where we generally assume that hazards arise accidentally or otherwise unintentionally. Airworthiness, a subset of safety, is defined as 'the ability to operate the air platform without significant hazards being presented'. Therefore, Cybersecurity for Airworthiness (CSA) characterises an intersection between aerospace cybersecurity and traditional safety (see Figure 1).

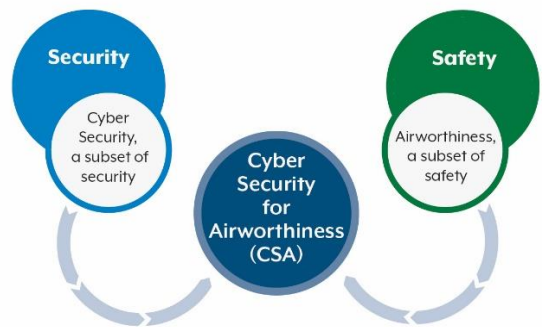

*Figure 1- Cybersecurity for Airworthiness*

---

[1] European Union Aviation Safety Agency [EASA] (2018) *RTCA DO-326A: Airworthiness Security Process Specification*, EASA.
[2] European Organisation for Civil Aviation Equipment [EUROCAE] (2014) *ED-202A – Airworthiness Security Process Specification*, EUROCAE.
[3] CSA is the UK Military Certification Specification for Airworthiness term for the UK Military delta - addition - to CS25.1319. This requires demonstration of compliance with the DO-326A set of standards combined with arguments against Joint Service Publication 440 - The Defence Manual of Security.

Continuous technological development has trended away from isolated systems and instead towards complex interconnected architectures, which bring with them a host of safety and security considerations; this is compounded further by the newer introduction of novel technologies such as Machine Learning (ML) models and other aspects of autonomy. The concept of CSA is best exemplified through looking at examples. Consider a cyber-attack by a threat actor which compromises the autopilot on a plane. The connected systems recognise an error condition and highlight the issue to the pilot, who takes over manual control - the system 'fails safe'. On-board safety has been maintained without compromising airworthiness. However, the threat actor successfully compromised the autopilot system.

*1.2 Problem Context*

The 'real-world' application of CSA is proving itself to be a complex task with many obstacles, and it is our understanding as authors that many organisations currently struggle with its implementation. We seek to understand why this is the case, and from there determine how best the obstacles may be cleared. The efficient and accurate implementation of CSA will provide wide-reaching benefits for all stakeholders across defence, commercial and civil/general air sectors.

In consideration of this problem, we leveraged our prior work in supporting CSA for multiple clients, as well as our experience in more traditional air safety and cyber assurance. Our involvement in key UK Ministry of Defence (MOD) working groups, and in support of Communities of Practice, was additionally useful in giving us inside knowledge of the barriers facing stakeholders. One output of our work is the development of processes, the Preliminary Aircraft Security Risk Assessment (PASRA), System Security Risk Assessment (SSRA) and related tools, for the UK MOD Defence Equipment and Support (DE&S) Airworthiness Team to enable teams to meet the intent of ED-202A, so that Chief Engineers may understand the levels of airworthiness risk that they hold. We have additional work underway to develop processes for communicating safety risks to operators.

We were able to produce a diagram of some of the key factors creating difficulty in applying CSA practices (see Figure 2). These factors will be explored in more depth in later sections.

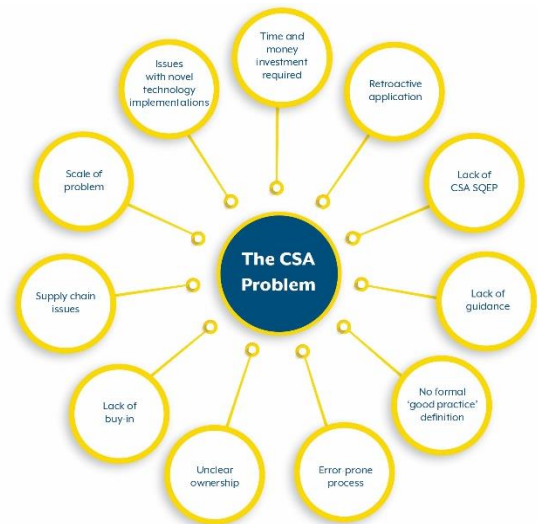

*Figure 2 – Factors limiting the ability to implement CSA.*

In previous research papers (see Section 2), we have focused primarily on the problem of 'retroactive application' – that is, the issue of attempting to apply new CSA frameworks to systems which were in some cases developed many years ago. Applying CSA practices is at its most beneficial when – and is often specifically intended to be – applied from the beginning of the design phase, therefore support organisations can struggle to effectively apply it to a platform already in service. The retroactive application problem also links somewhat to the SQEP and investment requirements, as they only compound the difficulty of attempting to apply standards such as EU 2023/203[4], DO-326A and ED-202A. We have previously recommended mitigating approaches to aid in tackling this problem;

---

[4] *Commission Implementing Regulation (EU) 2023/203* (2023). Available at: https://eur-lex.europa.eu/eli/reg_impl/2023/203

therefore, in this paper we will focus more on other factors, and the relationships between them, to discuss solutions which can lessen multiple burdens at once.

**2 - Factor Analysis**
Aside from retroactive application – which we have previously explored elsewhere – we considered each of the factors in turn, including how far they impact upon the ability to carry out CSA activities.

- Lack of CSA SQEP: There are significant differences between the realms of safety assurance and cyber assurance. In general terms, safety personnel consider both the intended and unintended outputs of a function, often – but not exclusively – classifying risks with values based on numerical calculations, whereas cyber personnel consider (mostly) intentional threats and usually perform no equivalent numerical scoring activity. There is necessity for CSA SQEP to understand the airworthiness, safety and cyber security domains. These skill sets tend to be contained in discreet functions and individuals in the current workforce, so there will be requirements for both training and a clear outline of required CSA competencies. It takes time and money to train practitioners and is further affected by a lack of training focused on CSA currently available. Compounding this is a lack of currently demonstrable experts, as CSA is a relatively new specialisation.
- Lack of regulatory guidance: There is little official guidance for how to satisfy CSA requirements, meaning that it can be hard to identify an effective process by which organisations can pursue its implementation. Whilst the amount of guidance is increasing, regulation will likely always be a step behind industry innovation. Furthermore, who is best placed to assume the role of regulator in this field, airworthiness or security? This question require future exploration as the practice continues to develop. For the UK MOD, this issue is compounded where there is a long-established security regulator, already implementing many of the requirements of the impending EASA CS-Information Security document.
- No formal 'good practice' definition: Due to the relevant recency of CSA as a concept, there are few examples of its application in practice. Of those that are available, there is difficulty in pinpointing a specific application of CSA which is considered as 'good practice'. This is exacerbated by the fact that standards are constantly developing resulting in a good practice against a standard today, becoming outdated and insufficient in just a short scale of time. The rate of novel technological development has a direct link to this factor as discussed further below.
- Scale of problem: In the case of modern air systems, architectures tend to be modular and deeply interconnected with one another. As the number of components to be considered in a modification or design increases, so does the amount of effort required for a sufficient analysis from the CSA perspective. In complex highly inter-connected aircraft – such as those we see in the military domain and particularly the ISTAR[5] role, the resource requirements can quickly approach insurmountable levels; context is therefore a key factor in the application of regulation and standards. The scale of the problem will be different between general aviation, where it will generally fall to the original equipment manufacturers (OEM), international carriers, regional carriers, and military operators.
- Error-prone process: Due to the scale of highly integrated architectures, a high level of effort is required to not only apply CSA to multiple complex systems, but also to ensure that it is done without overlooking anything or making costly mistakes. This issue of error likelihood scales exponentially with the level of architecture integration and complexity.
- Unclear Ownership: The priorities of safety and cybersecurity can come into conflict, and in the air domain safety generally takes precedence.       .
- Lack of understanding: From our experience, there appears to be an apparent lack of understanding from stakeholders as to the importance of cyber security, and the impact that cyber security can have on airworthiness and safety. This leads to it being poorly implemented, or not implemented at all in some cases, which often leads to significant ongoing airworthiness, safety, security, and operating risks.
- Supply chain issues: For Governments, the supply chains supporting armed forces carry an increased attractiveness to threat actors during the component's time in transit or maintenance, compared with commercial and general aviation. Assuring CSA requires assurance activities along the whole supply chain, presenting issues of in terms of complexity, understanding of CSA and a lack of obtainable evidence. In the case of in-service aircraft with service spanning decades, items of evidence may no longer exist in the supply chain. Whilst military operators may choose to accept a certain level of safety risk without applying security measures, this is generally not possible for civilian operators – meaning new security measures must be introduced. Some measures, such as securing the supply chain end-to-end, may not be feasible for civil operators.

---
[5] Intelligence, Surveillance, Target Acquisition, and Reconnaissance

- <u>Issues with novel technology implementations</u>: There is an increasing application of new or novel technologies in the air domain, such as Artificial Intelligence (AI), Machine Learning (ML) and Deep Learning (DL) components. Due to the novelty of these tools, and the 'black box' problem in the case of AI/ML, the process of assurance can be complicated, and requirements may be difficult to understand or entirely non-existent. With no regulation or legislation relating to the use of AI/ML, and standards being written and rewritten as the technology develops, this presents a challenge when ensuring or identifying safe and secure practice. It should be noted that the proposed EU legislation, *The Artificial Intelligence Act*, is currently draft at the time of writing and is yet to be formally introduced.
- <u>Time and money investment</u>: We have discussed several factors which stem from the complexity and seemingly wide ranging scope of the work required. This does not come without implications for workloads, timescales, and monetary investment requirements. Presently, with the rate of change being seen at time of writing, implementing CSA is a costly and intensive process and will likely remain one.

On deeper consideration, it seems that many of these factors are connected; for example, the time and money investment will scale with the amount of novel technology being used and the understanding of regulators. The use of novel or new technology can exacerbate the lack of guidance and or regulation, available examples of good practice, the lack of SQEP resource and an understanding of the regulations. A lack of understanding might stem from the conflicts between traditional safety and the newer CSA concepts, and from a lack of good practice examples.

Additionally, with the introduction of new or novel technology and the time it takes to apply the full extent of reasonable CSA practice, our idea of 'good enough' may have shifted. For a practical example, the use of quantum in overcoming traditional encryption; the rise of quantum computing now allows a threat actor to break commonly used encryption protocols in a time frame that makes this type of attack feasible for deployment against adversaries. At the end of developing and applying security measures, a platform or component may meet the required and relevant standard, but by that point, is this standard 'enough'? Figure 3 provides an illustration of influencing factors.

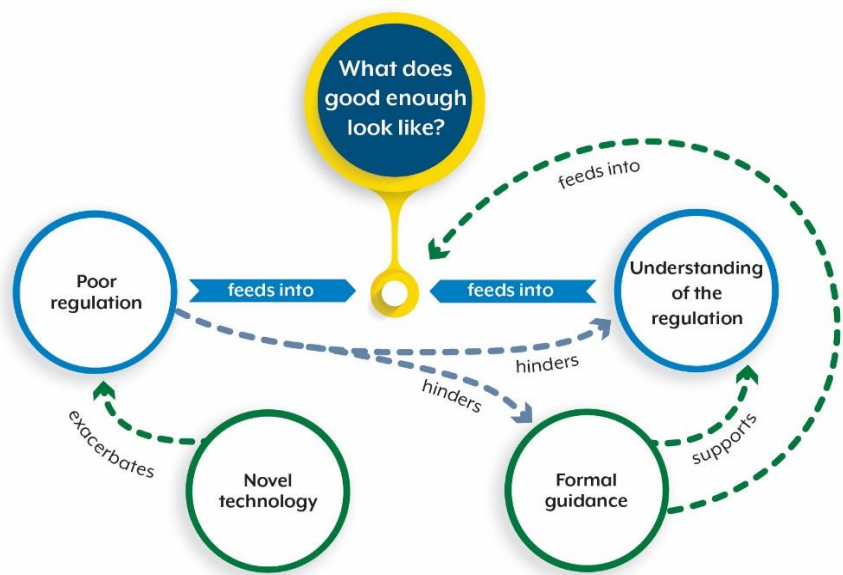

*Figure 3 - The issue of 'good enough' is a highly complex once, with interconnecting factors.*

Similarly, a solution to one problem may provide benefits to ease additional linking factors. This is one outcome explored in previous work, where our production of tools to provide direction and assistance in applying CSA discovered that one solution can mitigate many linked problems. Our prior work in this[6] and related topics[7] was predominantly aimed at addressing the retroactive application of Regulation or Standards, a lack of CSA SQEP and time and money resource required (see Figure 4).

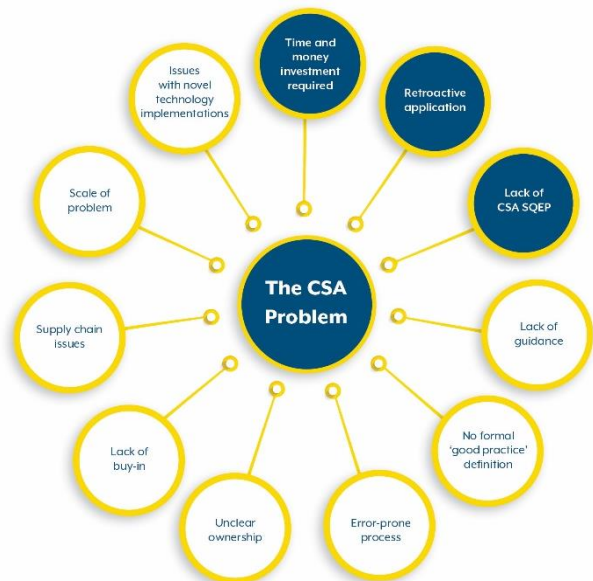

*Figure 4 - Issues handled in previous work.*

One aim of this paper is to propose measures that can assist in addressing the range of issues discussed, and build upon benefits identified in our previous work.

**3 - Proposed Way Forward**

Based on our experience in this field, we believe that any proposed solution should account for at least some of the following questions in order to be maximally effective:

- Does a smaller or less complex system require the same level of effort as a larger or more complex one? Should we apply the same level of regulation to a Grob G 120TP or Piper Cub as we would a Boeing 787 or Rafale?
- How does our approach compare to what other regulators are doing in regards to CSA? Do we put our platforms at greater airworthiness risk by not following suit?
- What does CSA SQEP look like? How should we train people in it?

We therefore believe that the first step to combating many of these problems is the introduction of comprehensive guidance for support organisations. This guidance should outline:

- Steps for meeting the 'baseline' requirement – note that our previous work in supporting the UK MOD included development of processes and templates to enable support organisations and teams in this task;
- Steps for applying best practice;
- What constitutes SQEP in the context of CSA.

The operation of aircraft by either commercial, general, or military operators face largely similar problems in current application of CSA practices, so it is likely that guidance and skills will be transferable to some extent. This would significantly ease multiple burdens currently being faced by support organisations:

---

[6] Ramsay, L and Warren, M, 2021, "Challenges of Cyber-Resilience in Aviation: An Implementation", *Safety-Critical Systems Club*.
[7] LeClair, B and Warren, M, 2021, "Maritime Cyber Safety – Results of Application to Real-World Systems", *International Maritime Conference 2022*.

- Lack of regulatory guidance: Agreed guidance would provide a path to standardised approaches that are applicable across a variety of anticipated use-cases, promoting efficiency[8]. It would also increase the accessibility of said guidance to stakeholders.
- No formal 'good practice' definition: This activity would include outlining good practice, which may have interoperability between the three types of aerospace spheres, either from OEMs, independent bodies, operators or regulators. This would allow support organisations to apply CSA more effectively across general, military, and commercial spheres. Additionally, scheduled opportunities to review the assurance activity and carrying out 'learning from experience' sessions to review 'good' or 'bad' actions
- Error-prone process: Issued guidance should aim to decrease the likelihood of application errors through sufficient stakeholder education or guidance on how to scale CSA assurance activities.
- Issues with novel technology implementations: Support in the assurance or application of novel or new technology through teaming of regulators, or by information sharing of beneficial approaches, could go some distance in increasing the total level of expertise surrounding these problems. The production of appropriate regulations based on identified good practice would go some distance in support of this.
- Lack of CSA SQEP: An agreement of what CSA 'looks like' in practice will generate clearer ideas on what CSA SQEP should be, allowing organisations to identify where and how to fill existing skills gaps in a time- and cost-effective manner. It is likely that many skills will be transferrable between sectors, presenting opportunities for collaborative SQEP building that more effectively addresses industry requirements.
- Lack of understanding: The wider business(es) should be supported in CSA education, and engineers will be able to use guidance to learn how to apply good practice effectively. This is especially important when considering that one of the key current issues is a lack of available education for stakeholders.
- Unclear ownership: Guidance should aim to provide clear roles and responsibilities for different activities, eliminating the problem of unclear ownership.
- Time and money investment required: The requirements of investing time and money should become clearer through the introduction of improved guidance, as it will give organisations a reasonable idea of the activities and education needs required to meet SQEP baselines.
- Retroactive application: Guidance should aim to support retroactive application, which currently forms the bulk of the work, as well as the implementation from initial design stages.

When comparing the potential impacts of this to the original set of factors, wide-reaching effects can be seen (see Figure 5).

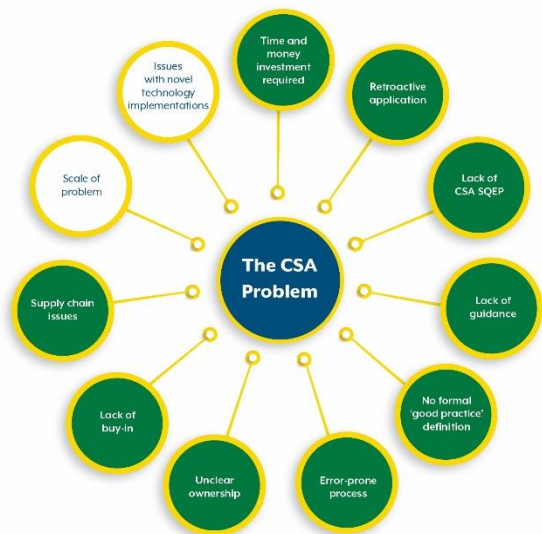

*Figure 5 - The green area shows aspects being addressed by clear guidance.*

---

[8] This does not mean, however, that there may not be additional considerations in certain contexts, and this should be assessed on a case-by-case basis.

This is not to say the task of introducing guidance is simple, rather there will be complexity and challenges in ensuring that proposed guidance is detailed enough to support stakeholders, provide a reasonable degree of efficacy, whilst also being flexible enough to support timely expansion and/or amendments in the face of rapid technological development. Additionally, it will be difficult to account for future technological developments and ensure that all bases are covered. We must, therefore, accept that guidance will always be lagging slightly behind technological development, but that we should not tolerate obsolescence of its guiding principles. There will be associated challenges for regulators in ensuring that they strike this balance, providing something that can quickly be expanded upon as required. This task would require facilitation of some means of regularly evaluating and aligning regulation to reflect current industry and technologies, which in turn carries implications for investments of time and work.

However, we believe the benefits to the aerospace sector would greatly outweigh the costs, should such an approach be implemented effectively. Such an activity for example, could potentially be the monitoring and evaluating the number of achieved CSA baselines over a period of differing system complexities. The activity would likely have continued benefits into the future.

The field of cyber security is especially turbulent, with new approaches, threats and mitigations emerging at a rate that is arguably impossible for a single regulatory body to match alone; this carries over to CSA. With more experts aligning to address key industry risks, more resource will be available to respond to developments in a timely manner. Additionally, facilitating these communications between standards and guidance bodies and regulators is likely to ease implementation of assurance activities beyond just those concerning CSA.

### 4 - Conclusions

In summary, we have presented and explained the various obstacles and barriers impacting application of CSA and have proposed a number of steps towards lowering the barrier. We believe the implementation of some of these would enable significant improvements over the current state of practice, where insufficient present approaches and a lack of unified sources of knowledge are creating a gap in airworthiness that has yet to be adequately quantified, yet alone addressed. The initial actions towards implementing our recommended changes would include communication links between relevant standards and guidance bodies in the air domain, and the implementation into support organisations. This could be done, for instance, through a community of practice that meets regularly to identify new ways of working, emerging technologies or upcoming changes to regulations or standards. Key initial activities should be identifying who the most relevant stakeholders are for attendance at these meetings, the objectives for the immediate term, and how results of discussions should be disseminated to industry. Ideally, this needs to be done across the entire air domain, irrespective of sphere.

There are still some further issues to be addressed by industry outside of this, namely how we can tackle the previously discussed issues regarding supply chains and the scale of application. There is clearly still space for significant innovation in the realm of CSA application; this will require continuous and targeted effort from industry experts.

### 5 - Bibliography

LeClair, B and Warren, M, 2021, "Maritime Cyber Safety – Results of Application to Real-World Systems", *International Maritime Conference 2022*.